\documentclass{iaus238}
\usepackage{graphicx}

\title[Black holes: from stars to galaxies] 
{Black holes: from stars to galaxies}

\author[I.~F. Mirabel]   
{I. F\'elix Mirabel\,\thanks{On leave from CEA, France.\hfill~}}

\affiliation{European Southern Observatory, Alonso de Cordova 3107, Santiago, Chile\\[\affilskip] 
email: fmirabel@eso.org}

\pubyear{2006}
\volume{238}  
\pagerange{001--999}
\date{to be defined}
\jname{Black Holes: from Stars to Galaxies -- across the Range of Masses}
\editors{V. Karas \& G. Matt}
\begin{document}

\maketitle

\begin{abstract}
While until recently they were often considered as exotic objects of
dubious existence, in the last decades there have been overwhelming
observational evidences for the presence of stellar mass black holes in
binary systems, supermassive black holes at the centers of galaxies, and
possibly, intermediate-mass black holes observed as ultraluminous X-ray
sources in nearby galaxies. Black holes are now widely accepted as real
physical entities that play an important role in several areas of modern
astrophysics. 

Here I review the concluding remarks of the IAU Sympposium No 238 on
Black Holes, with particular emphasis on the topical questions in this
area of research.

\end{abstract}

\firstsection 
\section{The aim of IAU Symposium No 238}
The interaction of black holes with their surroundings produces
analogous phenomena in AGN and stellar black hole binaries. The scales
of length and time of the phenomena are proportional to the mass of the
black hole, and the whole phenomenological diversity that takes place
around black holes can be described by the same physical concepts, however,
the observed phenomena exhibit enormous complexity (see figure~1).
Quasars and microquasars
can eject matter several times, whereas collapsars form jets only once.
When the jets point to the Earth these objects appear as microblazars,
blazars and gamma-ray bursts, respectively.
Synergy between research on stellar mass and supermassive black
holes has become essential for our understanding of the underlying
physics. 

\begin{figure}[tbh]
\includegraphics[width=\textwidth]{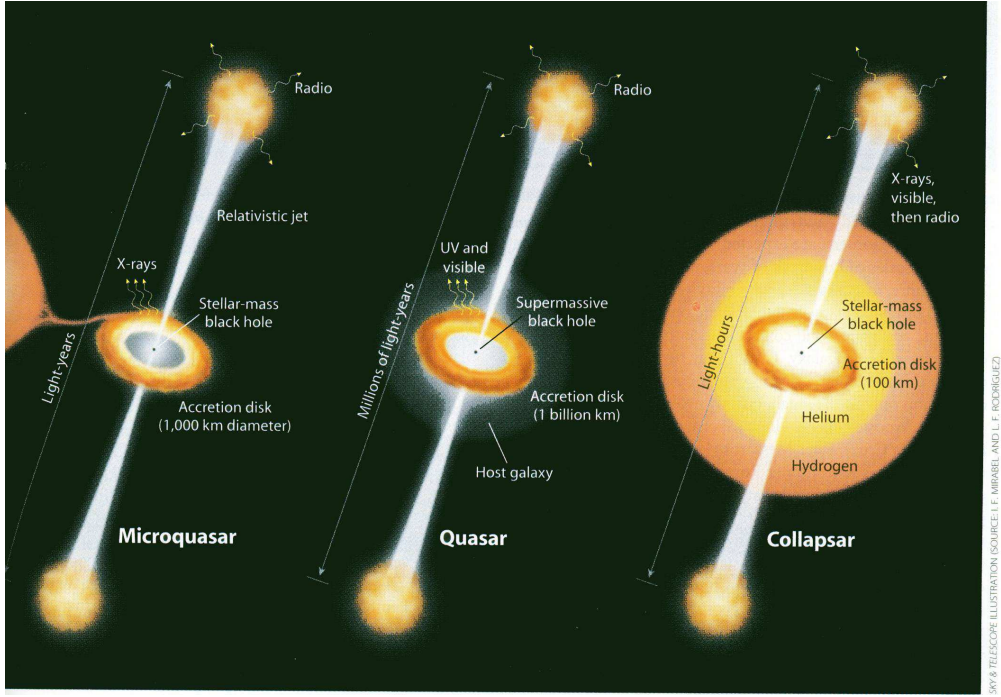}
\caption{The same physical mechanism can be responsible for the three
different types of objects: microquasars, quasars/AGN and massive stars that
collapse (``collapsars'') to form a black hole producing
Gamma-Ray-Bursts. Each one of these objects contains a black hole, an
accretion disk and relativistic particles jets.  Credit: Mirabel \&
Rodr\'\i{}guez (2002).}
\end{figure}

With this in mind, the aim of IAU Symposium No 238 has been to
discuss the relations between different types of astrophysical black
holes in a broader evolutionary context, bringing together astronomers
working on AGN with those working on compact stellar binaries. Several
groups that actively search for the elusive intermediate-mass black
holes have had an active participation.

\section{Stellar black holes}
Current physics suggests that compact objects in stellar binaries with
mass functions larger than 4 solar masses must be black holes. There are
about 20 known objects with such mass functions. They are believed to be
the tip of an iceberg, since it is estimated that in the Milky Way alone
there should be at least 1000 dormant black hole X-ray transients, while
the total number of stellar-mass black holes (isolated and in binaries)
could be as large as 100 million. The number statistics of known stellar
black holes is still very small and at present remain open the following
questions:  

Is the gap in the black hole mass function between 2.2 and 4.0 solar
masses real? If so, which is the physical reason?
Why in the Milky Way it has not been found a stellar black hole with
mass larger than 14 solar masses? Is this due to poor statistics or to
large mass losses in stellar winds by the metal rich progenitors in the
Milky Way? 

Kinematic studies of black hole binaries suggest that some stellar black
holes form associated to very energetic supernova explosions (see figure~2). 
In some cases, this kinematic evidence is reinforced by the chemical
composition of the donor star, when it contains elements produced in
supernova explosions. However, kinematic studies suggest that black
holes may also form by direct collapse, namely, in the dark (figure~3).
Therefore, it is an open question when stellar black holes form in
energetic supernovae and when by direct core collapse; more
specifically, whether the presence of an energetic  explosion depends or
not on the mass of the collapsing stellar core. In fact, the kinematics
of the microquasars Cygnus X-1 and GRS 1915+105, which contain black
holes of $\sim$ 10 and $\sim$ 14 solar masses, respectivelly, did not
receive strong kicks from natal energetic supernova explosions.

This question on the explosive or implosive black hole formation can be
approached by observations of nearby gamma-ray bursts of long duration,
which are believed to take place when stellar black holes are form by
core collapse of massive stars. Therefore, the question on whether all
collapsar GRBs are associated with supernovae of type Ia/b, or some are
not, is of topical interest for the understanding of the very last
phases of massive stellar evolution and black hole formation.  

\section{The supermassive black hole at the Galactic Center}

\begin{figure}[htb]
\begin{center}
\includegraphics[width=0.453\textwidth,bb=20 0 575 336,angle=0]{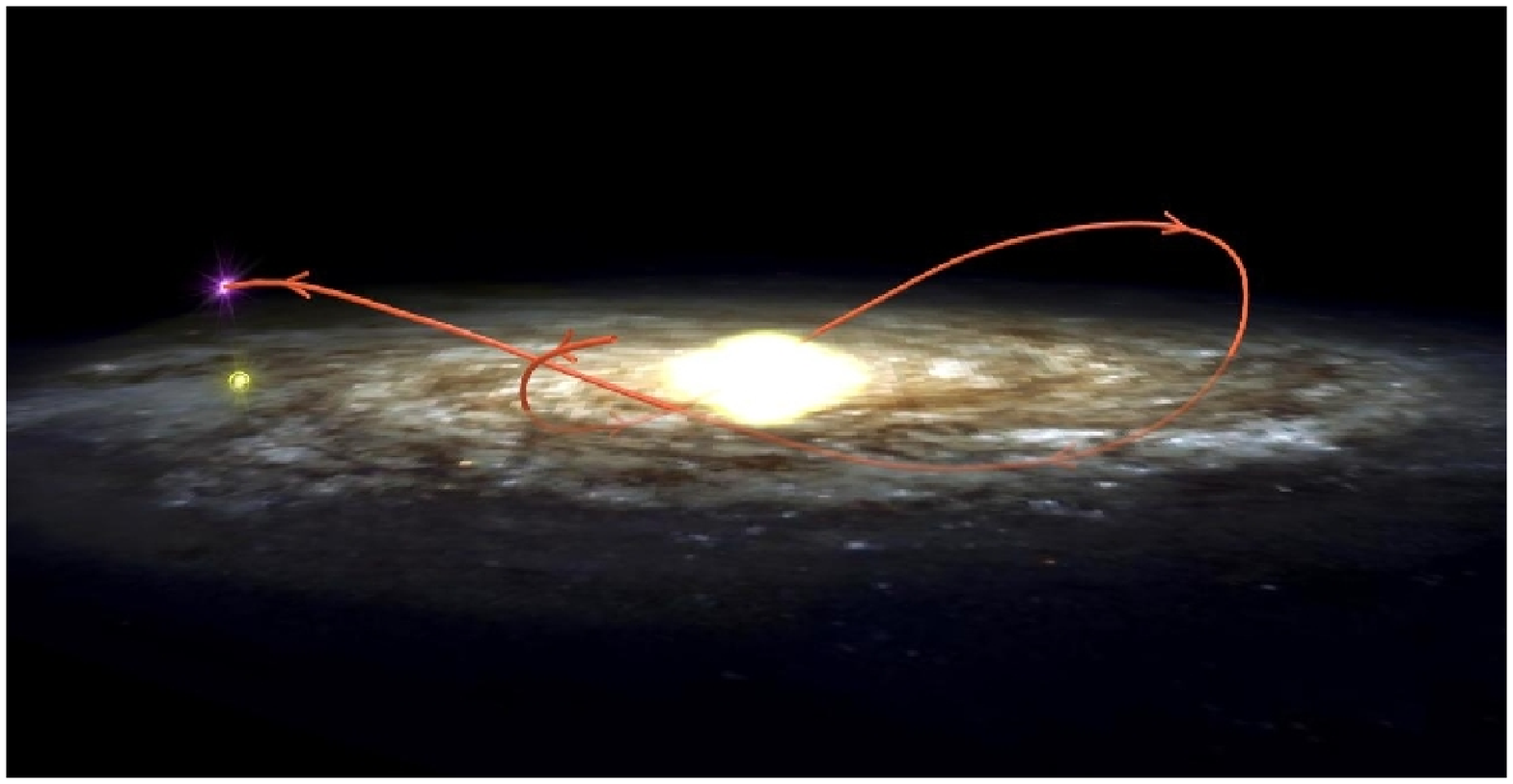}
\hfill
\includegraphics[width=0.537\textwidth,angle=0]{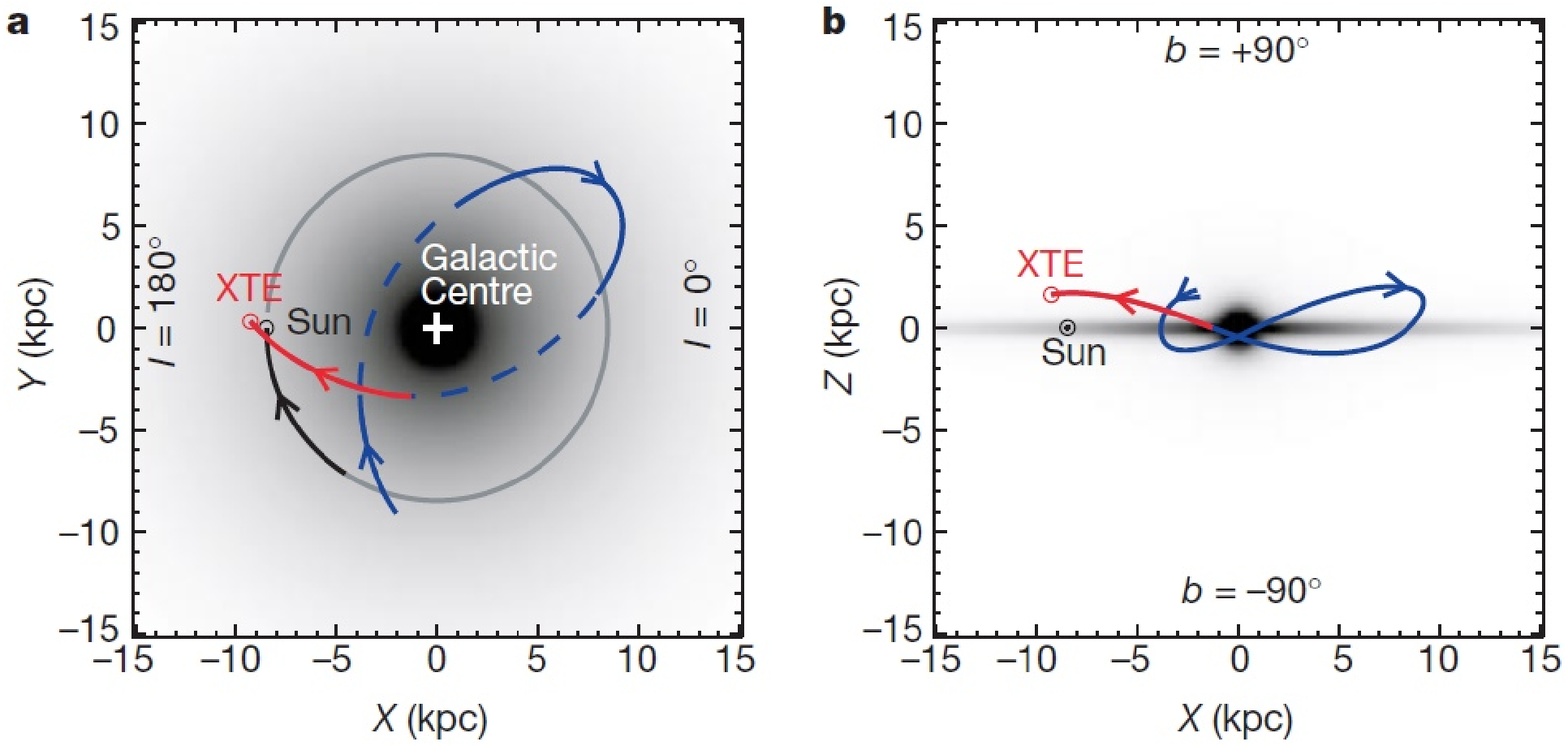}
\end{center}
\caption{Galactic trajectory of a black hole wandering in the Galactic
halo. This black hole may have been shoot out from the Galactic plane by
an energetic, natal supernova explosion. Credit: Mirabel \etal\ (2001).}
\end{figure}

\begin{figure}[tbh]
\includegraphics[width=0.47\textwidth]{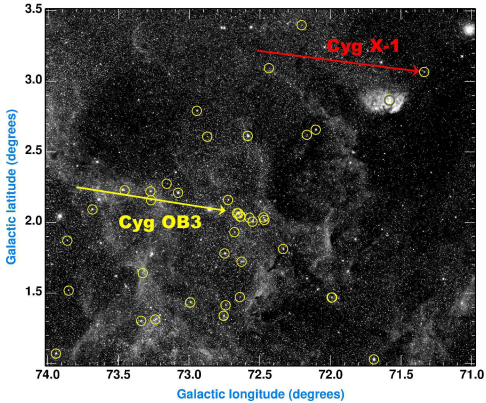}
\hfill
\parbox[b]{0.5\textwidth}{\caption{Kinematic evidence for the 
direct formation of the black hole
in Cygnus X-1. Optical image of the sky around the black hole binary
Cyg X-1 and the parent association of massive stars Cyg OB3. The red
arrow shows the motion in the sky of the radiocounterpart of Cyg X-1
for the past 0.5 millon years. The yellow arrow shows the average
Hipparcos motion of the massive stars of Cyg OB3 (circled in yellow) for
the past 0.5 millon years. Cyg X-1 moves along with the parent
association of massive stars indicating that the compact object did not
received an energetic kick from a natal supernova. Credit: Mirabel \&
Rodr\'\i{}gues (2003).\vspace*{3mm}}}
\end{figure}

Dynamics is the most direct method to determine the mass of
astrophysical compact objects, and therefore, the best evidence for the
existence of a black hole. The first unambigous dynamic evidence for a
supermassive black hole was found  following the motion of water masers
around the center of the galaxy NGC 4258. 

More robust evidence has been obtained by the motion of stars around
the dormant black hole of 3-4 million solar masses at the center of our
Galaxy. These stars seem to be distributed in two randomly inclined
disks of 0.04 pc and 0.5 pc radii.  The unexpected discovery of a
compact cluster of massive stars in the central parsec of our Galaxy
(see figure~4) poses new questions and may open new horizons for our
understanding of massive black hole formation and its relation with
massive star formation. This central cluster exhibits a flat mass
function and it is only 6 million years old. At present, it is not clear
how it got there. Perhaps it was formed in situ. How this could take
place under the strong tidal forces from the central black hole remains
a mystery.    

\section{Supermassive black holes in external galaxies}
Supermassive black holes are ubiquitous at the centers of galaxies.
Their mass is correlated with the mass of the host galaxy, and in
particular with that of the stellar bulge. This indicates that massive
black hole and host galaxy formation are tided up. 

\begin{figure}[tbh]
\includegraphics[width=0.45\textwidth,bb= 10 15 300 313,clip=true]{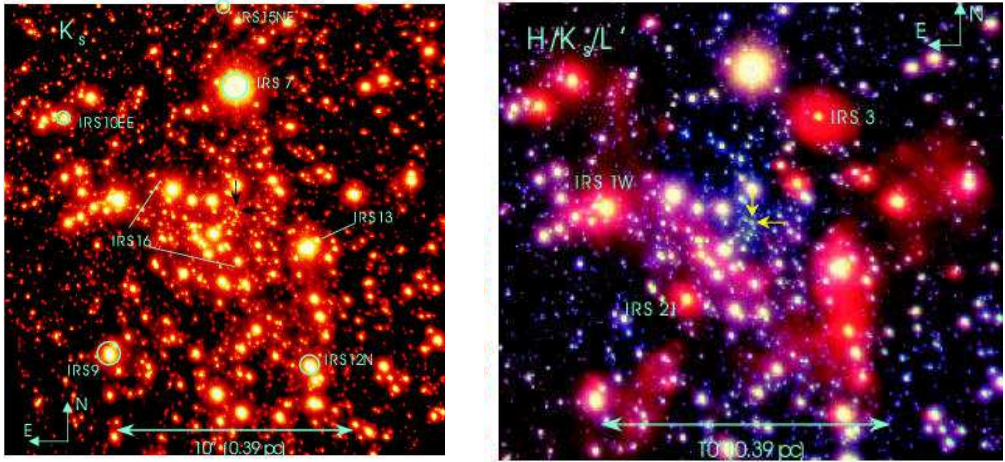}
\hfill
\includegraphics[width=0.51\textwidth,bb= 310 50 575 313,clip=true]{fig4.eps}
\caption{Left: The K-band near-infrared adaptive optics image of the 
central 20~arcsec of the Milky Way. Right: HKL colour composite image. 
Credit: Reinhard Genzel \etal\ (2003).}
\end{figure}

The most massive black holes are found at z $\geq$ 6.0 implying that
they assembled very early, in less than one billon years after the Big
Bang. On the other hand, supermassive black holes of lower mass have
formed more slowly by merging at z $\leq$ 3. The peak accretion rate
into black holes occurred rather late at z $\sim$ 0.7. Interestingly,
this redshift is similar to the redshift of the peak density of luminous
infrared galaxies at z   $\sim$ 0.8, which suggests the association of
maximum rate of accretion into black holes with dust enshrouded
starbursts in luminous infrared galaxies, which are known to be mergers
of gas-rich galaxies.

Although several models have been proposed, there is no general
concensus on how supermassive black holes were form. More specifically,
it is not known how could the most massive black holes in the universe
form so rapidly with no feedback. Did supermassive black holes form
before, after or coevally with the bulges?  

The kick velocity due to gravitational recoil in merging supermassive
black holes may displace the merged black hole from the dynamic center
of the host galaxies. For dwarf galaxies the estimated kick velocity is
larger than the typical escape velocities of 10-20 km/s, which may
result in the ejection to the intergalactic medium of naked, massive
black holes. How could we identify these runway black holes? 

\section{Black holes of intermediate mass}
The existence of black holes of intermediate mass remains an open
question. From the spectral properties of ultraluminous X-ray sources in
nearby galaxies it has been suggested that some of these sources contain
black holes of hundreds and perhaps thousands of solar masses. However,
no dynamic evidence from the motion of satellite objects around such
objects has been found. Furthermore, the luminosity function of black
hole binaries as well as the properties of the most luminous x-ray black
hole binaries in our Galaxy indicate that most of the super Eddington
sources observed in nearby galaxies are a natural extension of the
luminosity function of X-ray binaries, most of them being black holes
with masses smaller than 100 solar. It has been argued that the
hyper-accreting Galactic sources SS 433 and GRS 1915+105 seen from a
different angle would be classified as ultraluminous X-ray sources. 

If some ultraluminous x-ray sources are black holes of intermediate mass, one 
may ask why none has been identified in our Galaxy and/or the Magellanic Clouds, 
where the mass functions of compact objects can be determined. Black holes 
of intermediate mass may exist but are difficult to find.

\section{Correlations among black holes of all masses}

{a)}~The fundamental plane: At low accretion rates, correlations between
radio and x-ray luminosities are found for black holes across the whole
range of masses. At low rates of accretion most the radiation power is
dominated by synchrotron, compact, flat spectrum jets, from the x-rays
to radio waves. For black holes in the jet dominated state, universal
correlations between the x-ray luminosity, radio luminosity, and black
hole mass have been found. Using this black hole fundamental plane it is
expected that the masses of black holes could be determined from the
x-ray and radio luminosities.  

{b)}~Accretion--jet coupling: Because of their proximity and rapid
variability, microquasars have become the most adequate objects to study
the connection between instabilities in the accretion disks and the
genesis of relativistic jets. Figure~5 shows this connection as observed
in an interval of time of a few tens of minutes. This sequence has also
been observed in quasars, but on scales of a few years insted of tens of
minutes.

\begin{figure}[tbh]
\includegraphics[width=0.77\textwidth]{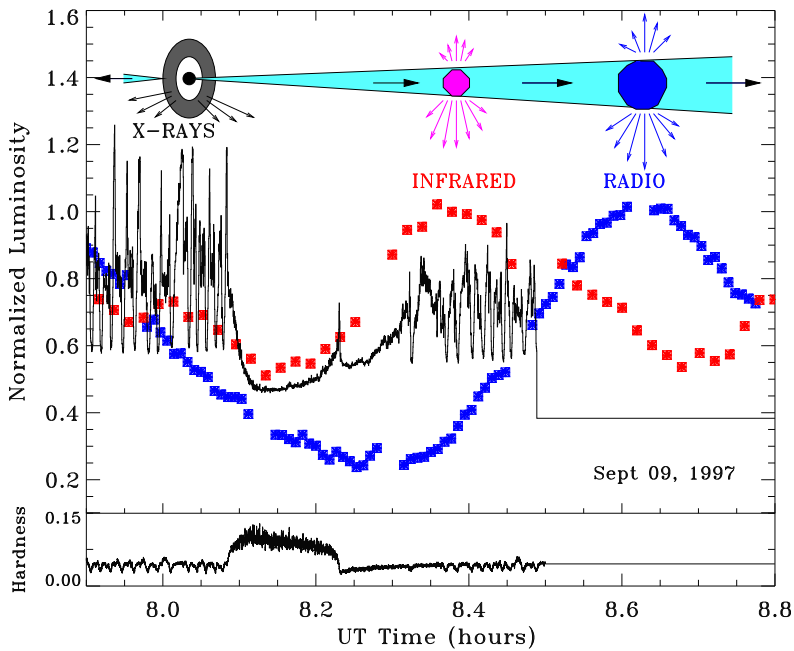}
\hfill
\parbox[b]{0.21\textwidth}{\caption{Real time observations in a 
microquasar of the connection
between instabilities in the accretion disk and the genesis of jets in
time scales of tens of minutes. Analogous disk--jet coupling has been
observed in quasars, but on time scales of years.  Credit: Mirabel \etal\ 
(1998).\newline\newline\newline}}
\end{figure}

{c)}~Massive outflows: Across all mass scales, black holes in high
luminous states exhibit X-ray absorption lines that reveal
sub-relativistic winds with mass outflows as large as 0.3 times the
accretion mass. Super-Eddington accretion makes massive outflows
inevitable because all the accreting energy cannot be radiated. 

In the case of black holes binaries the joint action of jets and massive
outflows heat and blow away the interstellar medium, as does the black
hole binary SS433 in the host nebula W50. X-ray observations with
Chandra and XMM-Newton show that AGN produce analogous impact on the
intergalactic medium, which has solved the long standing cooling flow
paradox.  

In this conference it was proposed that massive outflows may result from
the interaction of relativistic jets with warped accretion disks, in
both stellar and supermassive accreting black holes. Furthermore, it was
proposed that bended disks may explain several intriguing observational
results, such as the large numbers of obscured AGN, as well as the
re-direction of precessing jets, as in SS 433. The following questions
remain open: 

Are the relativistic jets in AGN and microquasars purely leptonic or
hadronic? 
Are the hadrons in the jets of SS433 a result of entrainment by the
relativistic jets in the massive winds? 
Is the nebula W50 that hosts SS433 the remnant of the natal supernova of
the compact object, or a blown away cavity by the super-winds, or both? 

{d)}~Quasi-periodic oscillations: Sgr A* accretes at low mass rates, and
quasi-periodic, polarized flares on scales of tens of minutes have now
been observed at X-rays, infrared, submillimeter and radio waves. At
longer waves the flares are polarized up to 10\%. From the time lag at
longer wavelengths it has been proposed that these flares could be
synchrotron self Compton emission from adiabatically expanding clouds.
It is unclear whether these expanding plasma clouds in Sgr A* are
rotating blobs in the accretion disk, or whether they are collimated
expanding jets as observed in microquasars. 

{e)}~Spin: Black holes are the simplest objects in the universe. They are
defined by only three parameters: the mass, the spin and the electric
charge. Because much of the radiation emerges within 6 gravitational
radii they provide a unique opportunity to probe gravity in the
strong-field regime. The radius of the ultimate stable orbit depends
from the spin, and knowing the mass, distance and inclination, it can be
derived the spin by measuring the X-ray flux and the temperature of the
accretion disk when the accretion disk is in the thermally dominated
state and at luminosities $\leq$ 0.3 Eddington. 

Another way to measure the spin is with skewed fluorescence iron lines.
Using both methods it has been claimed that some microquasars and AGN
host extreme Kerr black holes. The following questions remain
unanswered:
Would magnetic fields change completely the physical conditions in the
inner parts of accretion disks? 
Is there a general correlation between spin and jet power? 

\section{Historical and epistemological analogy between stellar astrophysics and black hole astrophysics}\label{sec:concl}

At present, black hole astrophysics is in an analogous situation as was
stellar astrophysics in the first decades of the XX century. At that
time, well before reaching the physical understanding of the interior of
stars and the way by which they produce and radiate their energy,
empirical correlations such as the HR diagram were found and used to
derive fundamental properties of the stars, such as their mass.
Analogous approaches are taking place in black hole astrophysics. Using
correlations among observables such as the radiated fluxes in x-rays and
radio waves, quasi-periodic oscillations, flickering frequencies, etc,
fundamental parameters that describe astrophysical black holes such as
the mass and spin of the black holes are being derived.

\bigskip\bigskip

\discuss{Phil Charles}{Could the apparent ``dark jet'' paradox of
SS433 be answered by the high inclination which causes obscuration of
most of the flux?}

\discuss{Felix Mirabel}{Yes, the accretion disc produces high opacity to
the X-rays and, as shown in the talk by Andrew King, the disc is
probably highly warped. The source could be very bright if seen from a
different direction.}

\discuss{Virginia Trimble}{Suppose there have never been those jets in
SS433, would there still be a supernova remnant there, or is the host
nebula blown mainly by the activity of central source?}

\discuss{Felix Mirabel}{This is an open
question. Clearly, the lateral extentions seen in the nebula W 50 that hosts 
SS433, have been blown away by the jets.}

\discuss{Gregory Beskin}{Is it possible to create jets without a black
hole in the centre of the disc, for example by a neutron star?}

\discuss{Felix Mirabel}{Yes indeed, there are jets in binaries with
confirmed neutron stars. Clear cases are Scorpius X-1 and Circinus
X-1.}

\discuss{Gloria Dubner}{Comment on the previous question by V.~Trimble:
There is a way to probe if the whole bubble was created by a compact
object. I believe that the whole nebula had been created by a supernova
remnant and its shape was then distorted by the action of SS433's jets.
The way to address this issue is by searching for spectral changes in
the radio emission -- the jets have different spectrum than the rest of
the bubble.}


\begin{thebibliography}{}

\bibitem[2003]{g03}Genzel~R., Sch\"odel~R., Ott~T. \etal\ 2003, ApJ, 594, 812

\bibitem[1998]{m98}Mirabel I.~F., Dhawan~V., Chaty~S. \etal\ 1998, A\&A, 330, L9

\bibitem[2001]{m01}Mirabel I.~F., Dhawan~V., Mignani R.~P., Rodrigues~I., Guglielmetti~F.
2001, Nature, 413, 139

\bibitem[2002]{m02a}Mirabel I.~F., Rodr\'\i{}guez L.~F. 2002, Sky \& Telescope, 103 (May), 32

\bibitem[2002]{m02b}Mirabel I.~F., Rodr\'\i{}guez L.~F. 2003, Science, 300. 1119 


\end{thebibliography}
\end{document}